# Development of a compact transmitter array for the acoustic neutrino detection calibration

S. Adrián-Martínez[*], M. Ardid, M.Bou-Cabo[*], G. Larosa, C.D. Llorens, J.A. Martínez-Mora

Institut d'Investigació per a la Gestió Integrada de les Zones Costaneres (IGIC) – Universitat Politècnica de València
C/ Paranimf 1, 46730 Gandia, València, SPAIN, e-mail: siladmar@ upv.es

[*] Multidark fellow

*Abstract*— **Parametric acoustic sources technique has been widely used in several fields of acoustics, especially in underwater acoustics with the aim to obtain very directive transducers. In this paper we present different studies and developments done during last years to develop a compact acoustic calibrator that allows emitting acoustic neutrino like signal with the goal to calibrate arrays of acoustic receiver sensors to detect ultra-high energy neutrinos.**

## I. Introduction

Recently undersea neutrino telescopes have become a real option to study astrophysical sources using neutrino detection. ANTARES [1-3] can be considered as the first complete undersea neutrino telescope installed and, it is being operated and taking data since 2008. This telescope is located at 40 km off the French coast near of Toulon at a depth of 2475m. It consists of 12 flexible strings that contain 25 storeys, each of them composed by 3 optical modules giving a global number of about 900 large optical sensors covering about 0.1 $km^2$ area and about 400 m of water column. Neutrino direction and energy reconstruction is done by detecting the Cherenkov light emitted by a muon that comes from neutrino interaction with matter with a very good time and position synchronization [4,5]. On the other hand, ANTARES also holds different kind of Earth-Sea science sensors which makes this facility not only an astrophysical telescope but also a very useful deep sea observatory. Besides all the scientific challenges pursued, it is also very remarkable the technological challenge that ANTARES represents.

On the other hand, and taking the experience acquired by NEMO and NESTOR projects and ANTARES neutrino telescope, the KM3NeT Collaboration has been formed to build and operate a new undersea neutrino telescope that will be at least 20 times greater than ANTARES. It has been designed and now it is in the preparatory phase for the construction. For this new neutrino telescope new solutions are necessary. Both experiments, ANTARES and KM3NeT consider acoustic detection as a possible and promising technique to cover the detection of Ultra High Energy (UHE) neutrinos, being possible to combine these two neutrino detection techniques for a hybrid underwater neutrino telescopes, especially considering that the optical neutrino techniques needs acoustic sensors as well for positioning purposes.

The first studies done in relation with the possibility of detecting ionizing particles by acoustic techniques were pointed out by Askarian in 1957. The thermo-acoustic model predicts that an acoustic signal can be produced from the interaction of a UHE neutrino in water. This interaction produces a particle cascade that deposes a high amount of energy in a relatively small volume of the medium, which form a heated volume that gives a measurable pressure signal. [6-8]. The simulations done predicts that 25% of the neutrino energy is deposed by a hadronic shower in a small almost cylindrical volume of a few cm of radius and several meters of length. Pressure signal generated has a bipolar shape in time and 'pancake' directivity, this means a flat disk propagation pattern that travel perpendicularly to the axis defined by the hadronic shower. As a reference example at 1 km distance, in direction perpendicular to a $10^{20}$ eV hadronic shower, it is expected that the acoustic pulse has about 0.2 Pa (peak-to-peak) in amplitude and about 50 µs in width. With respect to the directivity pattern the aperture angle of the pancake is expected to be of about 1°.

ANTARES has an acoustic detection system called AMADEUS that can be considered as a basic prototype to evaluate the feasibility of the neutrino acoustic detection technique. This system is a functional prototype array [9] composed by six acoustic storeys, three of them located on the Instrumentation Line and the other three on the 12 Detection Line, each storey containing six acoustic sensors. The system is operational and taking data. This system could be considered as a basic prototype to evaluate the feasibility of the acoustic detection technique. In spite of all sensors has been calibrated in the lab it would be desirable to dispose of a compact calibrator that "in situ" may be able to monitor the detection system, to train the system and tuning it in order to improve its performance to test and validate the technique and determining the reliability of the system [10]. We consider that for all this, it is critical having a compact transmitter able to mimic the signature of a UHE neutrino. In this paper the studies and development made to have such a transmitter based on the parametric acoustic sources techniques is presented.

## II. PARAMETRIC ACOUSTIC SOURCES

Acoustic parametric generation is a well know non-linear effect that was study first by Westervelt [11] at the 1960's. This technique has been studied quite extensively since then, being implemented in a lot of applications in the field of underwater acoustics, specifically with the aim to obtain very directive acoustic sources. The acoustic parametric effect occurs when two intense monochromatic beams with two close frequencies travel together through the medium (water for example). Under these conditions in the region of non-linear interaction, secondary harmonics appears with the sum, difference, and double spectral components. The application of the technique for the compact calibrator presents two additional difficulties: on one hand, it is a transient signal with broad frequency content, and the directivity has cylindrical symmetry. To deal with transient signals it is possible to generate a signal with 'special modulation' at a larger frequency in such a way that the pulse interacts with itself while travelling along the medium, providing the desired signal. In our case of study the desired signal would be a signal with bipolar shape in time. Theoretical and experimental studies [12] show that the shape of the secondary signal follows the second time derivative in time of envelop of the primary signal, following the equation

$$(1) \quad p(x,t) = \left(1 + \frac{B}{2A}\right) \frac{P^2 S}{16\pi \rho c^4 \alpha x} \frac{\partial^2}{\partial t^2} \left[ f\left(t - \frac{x}{c}\right) \right]^2$$

Parametric acoustic sources have some properties that result very interesting to be exploited in our acoustic calibrator:

- It is possible to obtain narrow directional patterns at small overall dimensions of primary transducer.
- The absence or low level of side lobes in a directional pattern on a difference frequency.
- Broad band of operating frequencies of radiated signals.
- Since the signal has to travel long distances, primary high-frequency signal will be absorbed.

For all these reasons, it seems that this technique could be interesting for the development of an acoustic compact calibrator.

## III. STUDIES TO EVALUATE THE TECHNIQUE

### a) Planar transducers

The first study to evaluate the parametric acoustic sources technique to be applied as emitter of acoustic neutrino like signal was to try to reproduce the bipolar shape of the signal. This study was done using planar transducers and it is described in [13]. The results in terms of the generation of the bipolar signal from the primary beam signal, the studies of the shape of the signal, the directivity patterns obtained, the evidence of the secondary non-linear beam generated at the medium and the check of the non-linear behavior with the amplitude of the primary beam were all coherent with the expectation from theory and demonstrated that the technique could be useful for the development of the compact acoustic calibrator able to mimic the signature of the UHE neutrino interaction.

### b) Cylindrical symmetry

Once we have been able to reproduce the shape of the desired signal using the parametric technique, the next step was, in one hand study with more detail the influence of 'modulated signal' used in emission with respect to the secondary beam generated, and on the other hand reproduce the 'pancake' pattern of emission desired using a single cylindrical transducer (Free Flooded Ring SX83 manufactured by Sensor Technology Ltd., Canada). This work is described in [14] and the results obtained agreed with the previous ones, but now dealing with the difficulties of the cylindrical symmetry and being able to obtain a 'pancake' directivity of a few degrees.

To verify the previous studies over longer distances, a new study using the same equipment and conditions but using a larger pool was performed in order to obtain the dependence of the pressure level as a function of the distance.
Figure 1 summarizes the results of this study by comparing the amplitude of the primary and secondary beams. A different behavior is observed between both, being a clear evidence of the generation of the secondary bipolar pulse by the parametric acoustic sources.

The studies for the evaluation of the parametric acoustic sources technique for the development of a transmitter able to mimic the acoustic signature of a UHE-neutrino interaction may be concluded as that it is possible to use the technique for this application. Moreover it presents advantages with respect other classical solutions: the use of a higher frequency in a linear phased array implies that fewer elements are needed in a shorter length having a more compact design, and thus, probably easier to install and deploy in undersea neutrino telescopes. A possible drawback of the system is that the parametric generation is not very efficient energetically, but since bipolar acoustic pulses from UHE-neutrino interactions are weak, they can be afforded having reasonable power levels of the primary beams. At this point, it is necessary to design the whole system array in order to be arranged in undersea neutrino telescopes or to be used in sea campaign calibrations.

There are two aspects that need to be dealt and solved, the mechanical design of the array and the electronics to drive the array of acoustic sensors.

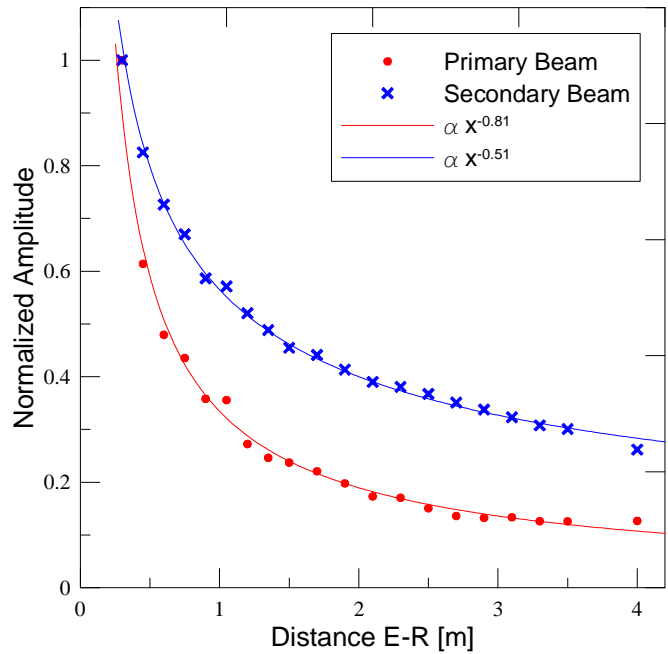

Figure 1. Amplitude of the primary and secondary signals as a function of the distance

IV. DESIGN AND TESTS OF THE COMPACT ACOUSTIC CALIBRATION SYSTEM

*a)   Array system*

•Design of the array

Once we have characterized the single cylindrical transducer for the generation of the bipolar pulse using parametric generation with the previous studies, we have all the input needed for the design of the array with this kind of transducers that will be able to generate the neutrino-like signal with the 'pancake' directivity with an aperture of about 1°. Figure 2 shows an example of the results of the calculations performed by summing the contributions of the different sensors for far distances at different angles. In this example, a linear array of 3 transducers with 20 cm separation from each other is enough to obtain an aperture of about 1°.

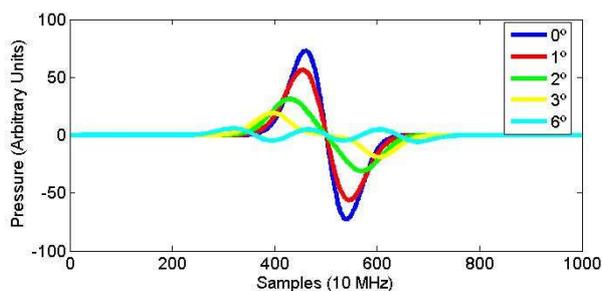

Figure 2. Pressure signal obtained at different angles for a three elements array.

•Experimental setup.

With the goal to reproduce the 'pancake' directivity, to cover long distances, and to improve the level of signal of non-linear beam generated at the medium, the array of three elements configuration has been proposed as possible solution It is composed by 3 Free Flooded Ring transducers, model SX83 manufactured by Sensor Technology Ltd., Canada. Each transducer has a diameter of 11.5 cm and 5 cm height. This transducer usually works around 10 kHz (the main resonance peak is at 10 kHz), but for our application we use a second peak resonance at about 400 kHz, which is the frequency used for the primary beam. The array developed for the tests has a separation between elements of 2cm, having the active part of the array a total height of 20 cm. The three elements are maintained in a linear array configuration by using three bars with mechanical holders as shown in Figure 3. The bars can help as well to hold the array in sea campaigns and to help to orientate it.

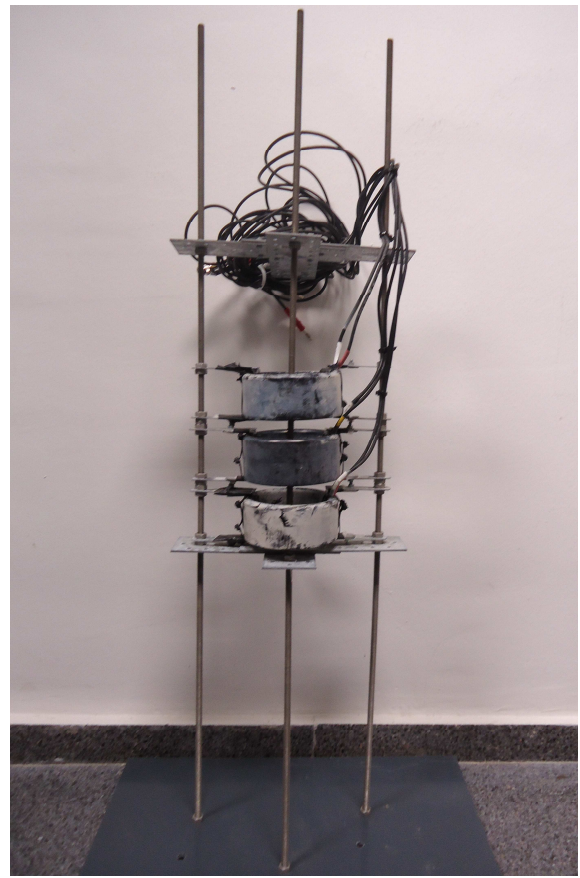

Figure 3. Picture of the array used for the tests.

The measurements for the array characterization have been made in an emitter-receiver configuration in a pool of 6.3 m length, 3.6 m width and 1.5 m depth. The array, which was 70 cm depth was used as emitter, and the receiver hydrophone used to measure the acoustic waveforms was a spherical omnidirectional transducer (model ITC-1042) connected to a 20 dB gain preamplifier (Reson CCA 1000). With this configuration the receiver presents an almost flat frequency response below 100 kHz with a sensitivity of about -180 dB, whereas it is 38 dB less sensitive at 400 kHz. The larger sensitivity at lower frequencies is very helpful to better observe the secondary primary beam. For these tests the emitter and receiver are aligned and positioned manually with cm accuracy, which is enough for our purposes. A picture of the pool during the tests can be seen in Figure 4.

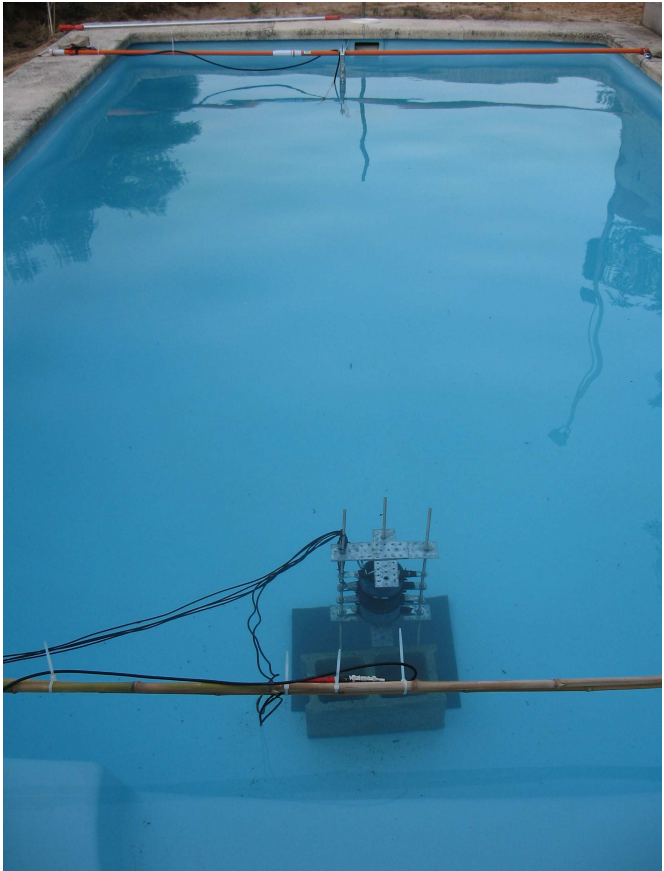

Figure 4. Picture of the pool during the data taking. The emitter array and the receiver (4.45 m apart) can be observed.

A DAQ system is used for emission and reception. To drive the emission, an arbitrary 14 bits waveform generator (National Instruments, PCI-5412) has been used with a sampling frequency of 10 MHz. This feeds a linear RF amplifier (ENI 1040L, 400W, +55 dB, Rochester, NY) used to amplify the emitted signal. In relation with reception, an 8 bit digitizer (National Instruments, PCI-5102) has been used with a sampling frequency of 20 MHz. Later, the data recorded is processed and different bandpass filters are applied to extract the primary and secondary beams signals and the relevant parameters.

- Results of the tests.

The results obtained for the array tests can be summarized in figures 5 and 6. In figure 5, an example of a received signal and the primary and secondary beams obtained after applying the bandpass filters (the secondary beam has been amplified by a factor 3 for a better observation) are shown. It is possible to see how the reproduction of the signal shape is achieved agreeing the results with our expectations from theory and previous observations. Figure 6 shows the directivity pattern measured at a longitudinal line defined by the axis of the array. The measurement has been made with a 4.45 m separation between the array and the receiver. It is worth to mention that for the secondary beam the FWHM measured with the array is about 7º (for a single element the FWHM was about 14º). Moreover, the pattern for the array should be a little more directive for far distances since the signals for the three elements will be better synchronized. Therefore, we can conclude that with this system is possible to have a 'pancake' directivity with an aperture of the 1º order.

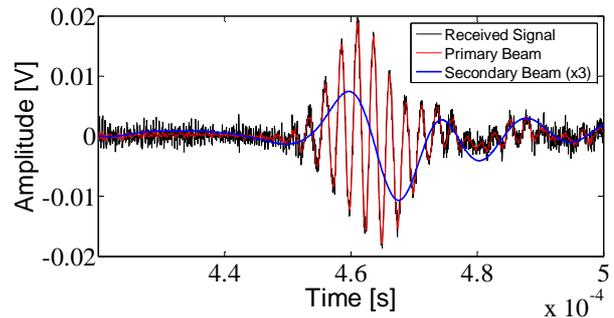

Figure 5. Example of a received signal and the primary and secondary beams obtained after applying the bandpass filters (the secondary beam has been amplified by a factor 3 for a better observation).

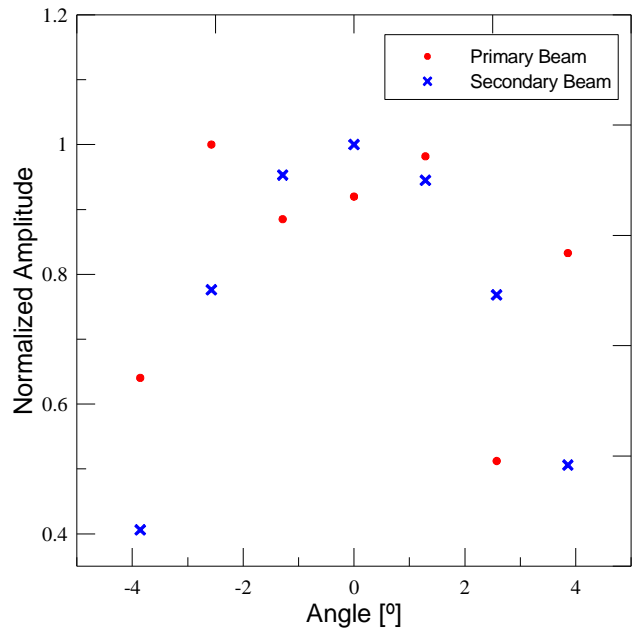

Figure 6. Directivity patterns of primary and secondary beam measured with the array.

- Calculations to cover long distances.

Taking into account the values of pressure calculated using the cylindrical array and in order to work within the underwater neutrino telescope infrastructure, it is useful to consider the extrapolation of the signal values for long distances.

Tables 1 and 2 presents the expected values for primary and secondary non-linear beam pressures, applying in one hand the attenuation due to radial divergence of the acoustic beam (1/r) and on the other hand the attenuation of medium (see table 3 to observe the conditions considered to calculate the attenuation value).

Table 1. Primary beam pressure at different distances

| Primary beam @1m | 60226 Pa |
|---|---|
| Pressure @ 1km | 0.003 Pa |
| Pressure @ 2km | 8.6·10-8 Pa |
| Pressure @ 3km | 3.1·10-12 Pa |
| Pressure @ 4km | 1.2·10-16 Pa |

Table 2. Secondary beam pressure at different distances

| Secondary beam @1m | 154 Pa |
|---|---|
| Pressure @ 1km | 0.1 Pa |
| Pressure @ 2km | 0.03Pa |
| Pressure @ 3km | 0.014Pa |
| Pressure @ 4km | 0.007Pa |

Table 3. Parameters used for the propagation

| Attenuation coefficients (Np/m) | α(25 kHz)=0.00042 |
|---|---|
|  | α(400 kHz)=0.00983 |
| T(ºC) | 13,2 |
| S(‰) | 38,5 |
| pH | 8,15 |
| depth (m) | 2200 |
| c(m/s) | 1541,7 |

It is possible to observe that despite the primary beam is dominant at short distances, the secondary beam is larger for long distances due to a lower absorption. These calculations have been made under very conservative assumptions since we have supposed that the secondary beam is attenuated by one over the distance, which is not the case at least for the measurements at short distances in the pool. We have not considered as well the possible improvements of the new electronics that are being developed, which will be adapted to the array, and more powerful signals are expected. Anyway, although the values are small in this conservative calculation the pressure obtained is similar to the pressure expected after a UHE-neutrino interaction, and therefore it is useful for this kind of calibration.

*b)   Electronics specifications*

With the goal of building an autonomous and optimized system, electronics associated to the acoustical array is under design and development process, producing a first prototype that is going to be tested. The novelty in the designing process of the electronic board is the use of the PWM (Pulse Width Modulation) technique to emit the necessary 'modulated signal' with the goal to obtain a secondary beam with the specifications desired. This technique has already been implemented by our research group in the electronics of the acoustic transceivers for positioning systems in underwater neutrino telescopes [15,16] Some of the advantages that offer this technique are:

- The system efficiency is improved because the system use a class D amplification, this means that the transistors are working on switching mode, suffering less power dissipation in terms of heat, and therefore offering a superior performance.
- Simplicity of design. Analog digital converters are not needed. It is possible to feed directly the amplifier with digital signal modulated by the PWM technique.
- It is not necessary to install large heat sinks at amplifier transistors, reducing the weight and volume of the electronics system.
- In waiting mode, the power amplifier has a minimum power at idle state that allows storing the energy for the next emission in the capacitor very fast and efficiently.

The electronic board is being designed using dsPIC33FJ256MC0710 [17], implementing the PWM under technology motorcontrol [18].

It is important to notice that working integrated into the infrastructure of an underwater neutrino telescope has usually constrains in power consumption. With the aim of test the feasibility of applying the PWM technique to emit our modulated signal, a simulation of the electronic board has been done. Pulse width modulated theory, works with the envelope of the desired signal used to feed the transducer. In our case this envelope is shown in figure 7.

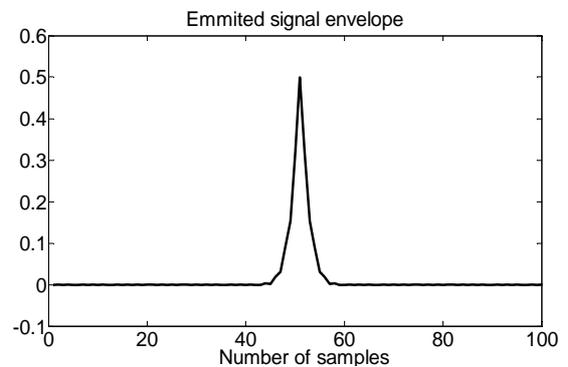

Figure 7. Envelope of the signal to be used for emission.

After applying PWM modulation it is possible to obtain the square signal that parameterize the signal before feeding the amplifier (figure 8). Finally, the result of the simulation shows the theoretical signal expected at the amplifier output (figure 9).

Seen the signals used in reference [14], it is clear that the result obtained by the simulation of the electronic board design implies that PWM is promising candidate technique in order to be implemented in the electronic board that will control the transducers array, being possible to drive the signal for feeding the transducer.

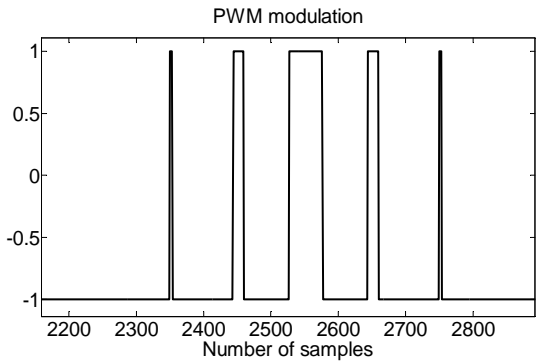

Figure 8. Square signal after applying PWM modulation.

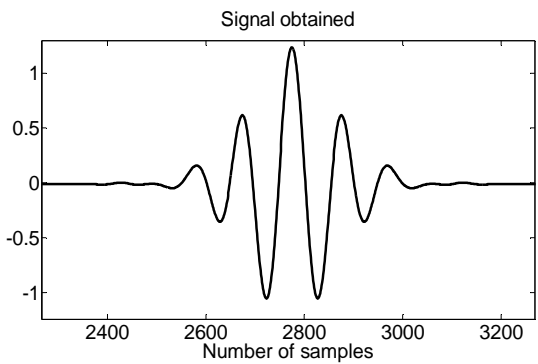

Figure 9. Expected signal at the amplifier output.

## V. Conclusions and future work

Considering all the results obtained in relation with the studies of the parametric acoustic sources, and with the first prototype of the array system, and following the results for the simulations, it seems clear that the solution proposed based in parametric acoustic sources could be considered as good candidate to generate the acoustic neutrino-like signals, achieving the reproduction of both specific characteristics the signal predicted by theory, bipolar shape in time, and 'pancake directivity'.

The near future work will consist in completing the the prototype of the compact transmitter and characterized it in the lab, and in situ. For the last part, a sea campaign is scheduled for the next September in the framework of the ANTARES neutrino telescope. The aim of the sea campaign will be to test the behavior of the transmitter using the AMADEUS system and try to perform some test to the AMADEUS system.

## Acknowledgments

This work has been supported by the Ministerio de Ciencia e Innovación (Spanish Government), project references FPA2009-13983-C02-02, ACI2009-1067, Consolider-Ingenio Multidark (CSD2009-00064). It has also being funded by Generalitat Valenciana, Prometeo/2009/26.